# The Plasma $\beta$ in quiet Sun Regions: Multi-Instrument View

Jenny M. Rodríguez-Gómez ,[1,2] Christoph Kuckein ,[3,4,5] Sergio J. Gonzalez Manrique,[6,3,4,7] Jonas Saqri,[8] Astrid Veronig,[8] Peter Gömöry[6] and Tatiana Podladchikova[9]

[1]*NASA Goddard Space Flight Center, Greenbelt, MD 20771, USA.*

[2]*The Catholic University of America, Washington, D.C. 20064, USA.*

[3]*Instituto de Astrofísica de Canarias 38205 C/ Vía Láctea, s/n, La Laguna, Tenerife, Spain.*

[4]*Departamento de Astrofísica, Universidad de La Laguna 38205, La Laguna, Tenerife, Spain.*

[5]*Max-Planck-Institut für Sonnensystemforschung, Justus-von-Liebig-Weg 3, 37077 Göttingen, Germany*

[6]*Astronomical Institute, Slovak Academy of Sciences, 059 60 Tatranska Lomnica, Slovak Republic*

[7]*Leibniz-Institut für Sonnenphysik (KIS), Schoeneckstr 6, Freiburg 79104, Germany.*

[8]*Institute of Physics & Kanzelhöhe Observatory, University of Graz. 8010 Graz, Universitätsplatz 5.*

[9]*Skolkovo Institute of Science and Technology, Territory of innovation center "Skolkovo", Bolshoy Boulevard 30, bld.1, Moscow 121205, Russia.*

## ABSTRACT

A joint campaign of several space-borne and ground-based observatories, such as the GREGOR solar telescope, the Extreme-ultraviolet Imaging Spectrometer (EIS), and the Interface Region Imaging Spectrograph (Hinode Observing Plan 381, 11-22 October 2019) was conducted to investigate the plasma $\beta$ in quiet sun regions. In this work, we focus on October 13, 17, and 19, 2019, to obtain the plasma $\beta$ at different heights through the solar atmosphere based on multi-height observational data. We obtained temperature, density and magnetic field estimates from the GREGOR High-resolution Fast Imager (HiFI), and Infrared Spectrograph (GRIS), IRIS, EIS and complementary data from SDO/AIA. Using observational data and models (e.g., FALC and PFSS), we determined the plasma $\beta$ in the photosphere, chromosphere, transition region and corona. The obtained plasma $\beta$ values lie inside the expected ranges through the solar atmosphere. However, at EIS and AIA coronal heights (from 1.03 $R_\odot$ to 1.20 $R_\odot$) plasma $\beta$ values appear in the limit defined by Gary (2001); such behavior was previously reported by Rodríguez Gómez et al. (2019). Additionally, we obtained the plasma $\beta$ in the solar photosphere at different optical depths from log $\tau$ = −1.0 to log $\tau$ = −2.0. These values decrease with optical depth. This work provides a complete picture of plasma $\beta$ in quiet sun regions through the solar atmosphere, which is a pre-requisite of a better understanding of the plasma dynamics at the base of the solar corona.

## 1. INTRODUCTION

The quiet sun is the region of the solar surface outside of sunspots, pores, and plages. This region exhibits impressive magnetic activity in different spatial scales and time scales, bringing large amounts of magnetic flux from the solar interior to the solar atmosphere. The quiet sun plays an important role in the solar atmosphere dynamics (Bellot Rubio & Orozco Suárez 2019; Faurobert 2017). In addition, variations in density and temperature above quiet sun regions can provide critical constrains on important open questions in solar physics (e.g., coronal heating and solar wind acceleration). However, the complexity of the solar atmosphere, observational challenges, and uncertainties in the derived parameters make this task difficult. A critical assessment of plasma diagnostic through the solar atmosphere is essential to understand the quiet sun's contribution on the solar dynamics at different heights.

An interesting feature in the quiet sun corona are coronal holes. They are considered the main source of the fast solar wind (Krieger et al. 1973) and a fundamental key in space weather (Linker et al. 2021). The predominantly "open" magnetic field allows plasma to escape (Altschuler et al. 1972). The predominant fraction of the "open" magnetic flux in coronal holes comes from long-lived magnetic elements with lifetimes >40 hrs (Hofmeister et al. 2019). Highresolution full-disk EUV filtergrams in several wavelengths sensitive to the 1-2 millions degree corona, e.g., 171, 193, 211 A from AIA aboard SDO (Lemen et al. 2012) are favorable to the observation and characterization of the coronal hole structure (e.g. Hofmeister et al. (2017)).

The study of plasma $\beta$ provides an essential description of the role of the kinetic and magnetic pressure in the solar atmosphere. The plasma $\beta$ distribution through the solar atmosphere was obtained mainly from models, e.g., Gary (2001) provides a model to calculate and delineate the limits of the plasma $\beta$ through the solar atmosphere. Recently, Rodríguez Gómez et al. (2019) presented a description of plasma $\beta$ through the solar cycles 23 and 24 using density and temperature estimations from the COronal DEnsity and Temperature (CODET) model (Rodríguez Gómez et al. 2018; Berdichevsky et al. 2022). However, the CODET model works mainly in the solar corona, between $\sim 1.14\ R_\odot$ and 2.42 $R_\odot$. Thus, from this approach, we cannot obtain plasma $\beta$ in the chromosphere and transition region. For this reason, in the present study we choose an observational approach. We obtained observing time via the H2020 SOLARNET program for the 1.5 m GREGOR solar telescope at the Observatorio del Teide, Tenerife, Spain. This observational campaign was supported by space-born observatories as well (HOP 381 from October 11 to 22, 2019). The main goal was to obtain density, temperature, and magnetic field to retrieve plasma $\beta$ through the solar atmosphere in the quiet sun regime.

In this paper, we obtain plasma $\beta$ through the solar atmosphere using a combination of observational data and models. Different techniques were used to retrieve density, temperature, and magnetic field in different atmospheric layers. We structure the paper as follows. The datasets are described in Section 2. In Section 3 We describe the methods to obtain density, temperature, and magnetic field at each layer using ground-based and satellite observations (e.g. photosphere, chromosphere, transition region, and corona). The main results are presented in Section 4. Finally, Section 5 contains a discussion and concluding remarks.

## 2. OBSERVATIONS

The dataset used in this study is composed of ground-based and satellite observations (Table 1). From 2019 October 10 to 18 the observations were centered at [0,0]″ and between October 19 to 22 at [0,500]″. The ground-based observations were obtained from the 1.5-meter GREGOR telescope (Schmidt et al. 2012). Specifically, we used the High-Resolution Fast Imager (HiFi; Denker et al. (2023); Kuckein et al. (2017)) and the GREGOR Infrared Spectrograph (GRIS; Collados et al. (2012)).

**Table 1.** Ground-based and satellite instruments, date, and layer observed on October 2019.

| Instrument | Date | Layer |
|---|---|---|
| The High-Resolution Fast Imager (HiFI) | Oct. 13 and 17 | Photosphere |
| The GREGOR Infrared Spectrograph (GRIS) | Oct. 13 and 17 | Photosphere and Chromosphere |
| The Interface Region Imaging Spectrograph (IRIS) | Oct. 19 | Transition Region |
| The Extreme-ultraviolet Imaging Spectrometer (EIS) | Oct. 17 and 19 | Corona |
| The Atmospheric Imaging Assembly (AIA) | Oct. 17 and 19 | Corona |

The HiFI observes the blue part of the spectrum (3850–5300 A). In this observational campaign, the blue continuum 450.5 $nm$, and G-Band 430.7 $nm$ filters were used. HiFI images have 2560 × 2160 pixels with a field of view (FOV) of 64.8″ × 54.6″. Thus, the image scale is about 0.025″/pixel. Dark and flat-field corrections were applied to all the images using the data reduction pipeline sTools (Kuckein et al. 2017). The G-Band and blue continuum images were co-aligned after the Speckel reconstruction, which was done using KISIP (Woger & von der Luhe 2008). HiFI consists of a synchronized two-camera system located in the blue wing of the 1.5-meter GREGOR telescope (Kleint et al. 2020; Schmidt et al. 2012). Each camera has a different filter so that simultaneous filtergrams in different wavelengths can be obtained. In our case, we used the G-band filter and a blue continuum filter. The 100 best images according to the Median-Filter-Gradient Similarity (MFGS; Deng et al. (2015)) were then selected and reconstructed using the speckle interferometry code KISIP (Wöger & von der Lühe 2008). We used HiFI images to obtain temperature estimates in the solar photosphere.

GRIS was used with a step size of the slit of $\sim 0.135''$/pixel. The spatial sampling along the slit was $\sim 0.136''$. It provides a maximum FOV of $60'' \times 40''$. The data was dark-current and flat-field corrected, and polarimetrically and wavelength calibrated (Collados 1999, 2003; Hofmann et al. 2012; Kuckein et al. 2012). In this study, we use data in the near-infrared He I 10830 A, which provides information on the chromospheric magnetic field and Si I 10827 A to infer plasma quantities as temperature, kinetic pressure, and magnetic field in the solar photosphere (González Manrique et al. 2020; Kuckein 2019). Weather conditions did not allow to have data during the complete observational campaign period. Thus, in this work, we used HiFI and GRIS observations on October 13 and 17, 2019.

The Hinode Observing Plan (HOP) 381 was executed from 11 October to 22 October 2019, delivering detailed data from the Extreme-ultraviolet Imaging Spectrometer (EIS: Culhane et al. (2007)) onboard Hinode and the Interface Region Imaging Spectrograph (IRIS: De Pontieu et al. (2014)).

IRIS provides simultaneously IRIS Slit Jaw Imager (SJI) and IRIS spectroscopic data. We used IRIS Slit Jaw Imager (SJI) in Si IV 1400 A as context images, and spectrograph data in the FUV (1332–1407 A) and NUV (2783–2835 A) ranges to obtain density and temperature estimates. Additionally, we used data from EIS observations of Fe XII 186.8 A, and Fe XII 195.12 A lines. EIS provides an excellent diagnostic of coronal plasma (Culhane et al. 2007; Kosugi et al. 2007; Young et al. 2007; Kayshap et al. 2015). To obtain additional information about the coronal plasma, we used full-disk filtergrams from the Atmospheric Imaging Assembly (AIA: (Lemen et al. 2012) onboard the Solar Dynamic Observatory (SDO: (Pesnell et al. 2012)) at 94 A, 131 A, 171 A, 193 A, 211 A and 335 A. The data were further processed using standard SolarSoft for IDL routines.

## 3. METHODS

Different methods to obtain temperature, density, and magnetic field are presented in the following subsections. All of them are needed to determine the plasma $\beta$ through the solar atmosphere. To retrieve the plasma $\beta$ values at each layer, we use the following relation:

$$\beta = \frac{2N_e k_B T_e}{B^2/8\pi}$$
(1)

where $N_e$ and $T_e$ correspond to the electron number density and temperature, $B$ is the magnetic field, and $k_B = 1.38 \times 10^{-16} erg \ K^{-1}$ is the Boltzmann constant.

### 3.1. *Photosphere*

We used two approaches to obtain the temperature in the solar photosphere using ground-based observations from HiFI and GRIS at the GREGOR solar telescope. The methodology described by de Boer et al. (1997) was applied to the HiFI datasets to obtain a temperature estimate. Assuming that the continuous radiation in the photosphere corresponds to that of a black body and according to Wien's approximation, we calculate

$$\frac{\ln(I_{Gband})}{\ln(I_{blue})} = \frac{\lambda_{blue}}{\lambda_{Gband}}$$
(2)

To the speckle reconstructed HiFI images (de Boer et al. 1997). $I_{Gband}$ and $I_{blue}$ correspond to the HiFI intensity of G-band and blue continuum at 430.7 *nm* and 450.5 *nm*, respectively. Using Wien's approximation is possible to obtain a temperature estimate of the blue continuum and G-band at the solar photosphere as follows (with wavelength $\lambda$ is given in nm and temperature in K):

$$T_{blue} = \frac{2.897 \times 10^6}{\lambda_{blue}} \, [K] \tag{3}$$

$$T_{Gband} = \frac{2.897 \times 10^6}{\lambda_{Gband}} \, [K] \tag{4}$$

$\lambda_{blue}$ and $\lambda_{Gband}$ can be obtained from Equations 3 and 4, and replaced in Equation 2. To obtain the temperature in function of the intensity obtained from the speckle reconstructed HiFI images.

The Stokes profiles acquired with GRIS were normalized to the continuum, which was determined by averaging an area of quiet-Sun of 145 × 10 pixels within the FOV. The wavelength array was determined on an absolute scale following the steps described in the appendices of Martínez Pillet et al. (1997) and Kuckein et al. (2012). To that end, two telluric lines within the spectral range served as a reference to extract the spectral sampling. The wavelength array was then corrected for orbital motions, solar rotation, and gravitational redshift. The wavelengths at rest were taken from the National Institute of Standards and Technology (NIST; Kramida et al. (2023)). As part of the pre-processing of the data, a 3 × 3 binning in the spatial dimension, and a 3-pixel binning in the spectral dimension, was carried out to enhance the signal-to-noise ratio. The resulting pixel size is about 0.405 arcsec$^2$ and the spectral sampling is 0.0539 A/pixel.

The physical parameters were inferred from the Stokes profiles of the Si I 10827 A line using the Stokes Inversions based on Response functions (SIR, Ruiz Cobo & del Toro Iniesta (1992)) code, which assumes local thermodynamic equilibrium (LTE) and hydrostatic equilibrium. The code solves the radiative transfer equation iteratively. SIR operates in logarithmic units of the optical depth at 5000 A to provide height-dependent physical parameters. We used the upgraded parallelized version of SIR (Gafeira et al. 2021). Previous studies have noted a significant improvement of the fits to the Stokes profiles when using different initial atmospheres (e.g., Quintero Noda et al. (2014); Kuckein et al. (2021)). Therefore, we inverted each pixel eight times, using each time different initial-guess atmospheres out of eight pre-selected atmospheres spanning between log $\tau$ = 1.0 and log $\tau$ = −3.8. The best fit is kept according to the lowest $\chi^2$ value, which is the sum of the squared differences between the observed binned Stokes profiles, and the synthesized profiles from the SIR code. The macroturbulence was left as a height-independent free parameter and the amount of diffuse straylight was fixed at 10%. The abundances were taken from Asplund et al. (2009). We used the following amount of nodes for the photospheric SIR inversions: up to four nodes for the temperature, one node for microturbulence and azimuth, up to two nodes for the magnetic field strength and its inclination, and up to three nodes for the LOS velocity.

### 3.2. *Chromosphere*

The He I 10830 A triplet has the potential of measuring chromospheric magnetic fields (Harvey & Hall 1971). The formation height lies just below the transition region (upper chromosphere) between 1500 km and 2000 km (Lagg 2007) at a temperature of about 7000 K (Tlatov 2003; Avrett et al. 1994a). The He I 10830 A lines are relatively weak in comparison to other chromospheric lines, but represent a powerful tool for understanding chromospheric processes (Huang et al. 2020). In general, the Helium 10830 A multiplet provides a good diagnostic for exploring the dynamic and magnetic properties of plasma in the solar chromosphere. Because their polarization is sensitive to the presence of atomic level polarization and the joint action of the Hanle and Zeeman effects, makes these lines sensitive to different magnetic field regimes (Centeno et al. 2008). We employed the HAnle and ZEeman Light v2.0.1 (HAZEL 2[1]) code for inverting the He I triplet. HAZEL 2 incorporates the effects of atomic level polarization, the Paschen-Back, Hanle, and Zeeman effects for the He I triplet. This code employed a cloud model with constant atmospheric parameters within a slab above the solar surface. To fit the Stokes profiles, we used a two-component model for each pixel. One component represented the inferred atmosphere, while the other accounted for stray light. The contamination related to stray light can occur due to varying seeing conditions, diffraction effects and telescope optics. We assumed a constant stray-light profile for all locations and times throughout the temporal series, like the SIR inversions described in Section 3.1. The inversions operated under the assumption that the observed radiation from a resolution element results from the combined contribution of the solar atmosphere at that position and spurious stray light.

---

[1] The inversion code used was HAZEL 2 and the current version can be found at https://github.com/aasensio/hazel2

HAZEL 2 is similar to a Milne-Eddington inversion code, meaning there is no height stratification. Non-LTE processes are not used here. The same Stokes profiles acquired with GRIS and data pre-processing described in section 3.1, including the normalization, wavelength calibration, and the 3×3 binning in the spatial dimension, together with a 3-pixel binning in the spectral dimension were used. The profiles observed clearly show only a single component in the He I 10830 multiplet. Thus, they were inverted with a single chromospheric component to determine the vector magnetic field, optical depths, velocities among other physical variables. To ensure better inversion and fitting results we decided to use the whole 10830 A region. This means we took into account also the Si I and the closest telluric line to the He I triplet. We selected the best atmospheric models that correlate with our observations. The models enclose the photosphere, assuming LTE, that produces the Si I 10827 A line, one single chromosphere in the He I 10830 A multiplet, and a simple Voigt function that fits the telluric line around the 10832 A triplet. The photospheric Si I 10827 A spectral line was included to ensure a better fit of the wavelength region, but the results of the LTE inversions of these atmospheric levels were not used in our analysis. The telluric line is always added to guarantee a good fit in the cases where the He I multiplet blends with the telluric contamination due to high redshift. The inversions were implemented for three cycles. The first cycle fits the Stokes I profiles of the 10830 A region. The magnetic field is not inverted during the first cycle. In the second and the third cycle, all the Stokes parameters are fitted both in the chromosphere and the photosphere. We found out that by giving higher weights during the second and third cycle to the Stokes Q, U, and V profiles, the inversion gives better fitting result since the signal of those Stokes parameters is low in many of the pixels.

### 3.3. *Transition Region*

We use IRIS spectroscopic level 2 data to measure the intensity $I_{obs}$ (in data number (DN)) of the Transition Region lines in the observed spectra. These datasets were obtained from level 0 data after flat-field, geometry calibration and dark current subtraction (IRIS software notes [2]). Additionally, the Solarsoft routine *despik.pro* was applied to remove cosmic ray hits. Calibration of the wavelength scale was performed (Polito et al. 2015, 2016b), and the intensities were converted to physical units. This procedure is necessary to estimate the electron number density $N_e$ of the emitting plasma.

The density diagnostic is based on the intensity ratio of two lines from the same ion. This method can be applied to spectral lines from the OIV and SIV ions, which are formed at close values of temperature and density, in a constant-pressure plasma. The OIV density-sensitive intensity ratios used in this work are

$$R_1 = \frac{I_{OIV}(1699.77\,A)}{I_{OIV}(1401.16\,A)} \tag{5}$$

The ratio $R_1$ is obtained by using the atomic data and it provides an useful density diagnostic in the transition region (Polito et al. 2016a).

### 3.4. *Corona*

To obtain estimates of density and temperature in the solar corona, we used data from EIS and AIA. EIS intensities from Fe ions were obtained in order to retrieve density estimates. Additionally, EUV data from SDO/AIA were used to perform Differential Emission Measure (DEM) analysis, and to obtain density and temperature in the solar corona. A brief description of the techniques applied to each kind of data is presented in the following.

---

[2] https://iris.lmsal.com/documents.html

[3] https://www.chiantidatabase.org/

### 3.4.1. *Extreme-ultraviolet Imaging Spectrometer (Hinode/EIS)*

Hinode/EIS allows performing plasma diagnostics based on the ratio of two spectral lines, one of which is excited from the ground level and the other from a metastable level, the population of which depends strongly on the local electron density (Young et al. 2007; Graham et al. 2011). EIS has a high sensitivity in the range 185 – 205 A, this interval provides an excellent density diagnostic from Fe XII and Fe XIII lines. Density estimates from EIS obtained using the Fe XII and Fe XIII have been presented in Brooks et al. (2007); Watanabe et al. (2007); Doschek et al. (2007); Dere et al. (2007); Young et al. (2009) and references therein. They demonstrated the high quality of the density estimates using EIS. The EIS data were calibrated from the level-0 FITS files using the EIS PREP routine, which is available in the *solarsoft* IDL distribution. A line fitting procedure was applied to obtain density diagnostic fitting the emission lines (Fe XII 186 A and 195 A) with Gaussians. The IDL routine EIS GETWINDATA automatically fit, the calibrated line intensities from a spectral line window. The spectrum from each image pixel is fitted, with data points weighted with $1\sigma$ error bars calculated by EIS PREP (Young et al. 2009). To convert a measured line ratio value to a density value, we used the CHIANTI atomic database 8.0 [3] (Del Zanna et al. 2015). CHIANTI contains atomic data for calculating line ratios as a function of density.

### 3.4.2. *Differential Emission Measure (DEM) analysis*

We used the six coronal EUV channels of the AIA Instrument (94A, 131A, 171A, 193A, 211A and 335A) to obtain density and temperature estimates in the solar corona through the DEM method. For optical thin emission from plasma in thermodynamic equilibrium, the temperature distribution of the contributing plasma in the line-of-sight (LOS) $h$ is described by the differential emission measure defined as

$$DEM(T) = N_e(T)^2 \frac{dh}{dT}$$

(6)

where $N_e$ is the number density dependent on the temperature $T$ along the LOS. The measured intensity $I_\lambda(T)$ for a given AIA wavelength is related to the DEM as

$$I_\lambda(T) = \int_T R_\lambda(T) DEM(T) dT$$

(7)

where $R_\lambda$ corresponds to the response function of each filter. We applied the regularized inversion technique developed by Hannah & Kontar (2012) to reconstruct the DEM curve for each binned pixel. The total emission measure was obtained by integrating the DEM over the full temperature range. The mean temperature can be estimated by the emission weighted temperature. Assuming a filling factor of unity, and estimating the the column height ($h$) of the emitting plasma along LOS, by the hydrostatic scale height, we derive $h$ as

$$h = \frac{K_B T_e}{\mu m_H g_\odot}$$

(8)

Where $\mu$ is the molecular weight, $m_H$ is the hydrogen mass and $g_\odot$ the solar gravitation (Aschwanden 2005; Aschwanden & Nitta 2000). To estimate the plasma density ($N_e$), we used the hydrostatic scale height and the emission (EM) obtained from DEM calculation (Saqri et al. 2020):

$$N_e = \sqrt{\frac{EM}{h}}$$

(9)

## 4. RESULTS

In this section, we presented our main results, namely the temperature, density, and magnetic field estimations as well as plasma $\beta$ at each layer using the different techniques applied to ground-based and satellite datasets described above.

### 4.1. *Photosphere*

Two different approaches are presented in this section. Temperature estimations using HiFI datasets and temperature, kinetic pressure and magnetic field estimates from the GRIS datasets. First, we calculated the temperature using speckle-reconstructed HiFI data from October 13 and 17, 2019. The temperature was obtained using Equations 2, 3 and 4 in the blue continuum and G-Band wavelengths. The percentage of the error was calculated as

$$err = \left| \frac{theo - calc}{calc} \right| \times 100\% \tag{10}$$

where *calc* are the values obtained using HiFI data and *theo* corresponds to theoretical temperature values for the blue continuum (450.55 *nm*) and the G-band (430.70 *nm*) from Equations 3 and 4. Theoretical values correspond to

$T_{blue}$ = 6430 *K* and $T_{Gband}$ = 6726 *K*. On October 13 at 08:55:15 UT, we used the complete FOV (64.8″ × 54.6″),

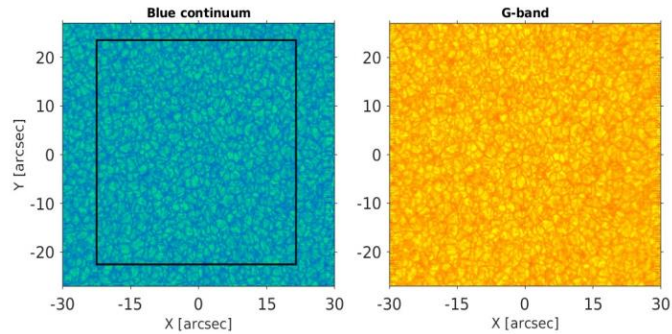

**Figure 1.** HiFI Speckle-reconstructed GREGOR images on Oct. 13, 2019 at 09:04:51 UT in the blue continuum (left), G-band (right) wavelengths. The black square shows the region over which the temperature estimate was obtained.

and the mean temperature obtained for the G-band is 6866 *K* with an error of 2.0%. At 09:04:51 UT we used a small FOV (45″ × 45″), and we obtained a mean temperature in the blue continuum of 6394 *K* with an error of 0.5%. On October 17 at 09:35:41 UT we used a small FOV (51.5″ × 51.5″), the mean temperature in G-band corresponds to 6783 *K* with an error of 0.8% (Figure 1). The reason is that the adaptive optics system from GREGOR corrects better the center of the image.

We used HMI/SDO data for the same FOVs on October 13 and 17, 2019. The quiet sun regimes on magnetic field were defined as $B = 5 - 50$ *G* (Rodríguez Gómez et al. 2019). Specifically, this threshold describe the small-scale magnetic elements in the photosphere. Density values related to the quiet sun were obtained from the FALC model. We selected models 1000, 1001, 1002 for quiet Sun-internetwork, dark quiet Sun internetwork, and quiet Sun network lane (Fontenla et al. 2011). We used the total density as $n = N_e + N_p + N_H$, where $N_e$ is the electron density, $N_p$ proton density and $N_H$ correspond to the neutral hydrogen density [$cm^{-3}$]. All of these values were used to calculate the plasma $\beta$ using Eq.1 and summarized in Table 2.

**Table 2.** Plasma $\beta$ in the solar photosphere using HiFI data, the mean magnetic field value from HMI/SDO, and the total density from FALC model.

| Day     | B[G] | Plasma $\beta$ |
|---------|------|----------------|
| Oct. 13 | 8.16 | 3.28           |
| Oct. 17 | 8.68 | 2.41           |

Additionally, we used the SIR code in the observed GRIS spectra to retrieve plasma parameters and magnetic field in the solar photosphere. The SIR code provides estimates of gas pressure, electronic pressure, LOS velocity, and magnetic field in the solar photosphere at different heights from the Si I 10827 A line. SIR assumes local thermodynamic equilibrium (LTE) and hydrostatic equilibrium. In this analysis three different optical depths were selected log $\tau$ = −1.0,−1.5 and −2.0. The optical depth log $\tau$ = −1 corresponds to the region where the spectral lines are most sensitive to changes in plasma parameters (Thonhofer et al. 2012; Cabrera Solana et al. 2005). Figure 2 shows temperature ($T$ [$K$]), kinetic pressure ($P_k$ [$dyn\ cm^{-2}$]), magnetic field ($B$ [$G$]) from SIR code inversions at log $\tau$ = −1.0 on October 13 from 08:55:21 to 09:00:38 UT (a) and October 17 from 09:48:45 to 09:59:16 UT (b). On both days the center of the observations is [0",0"]. Additionally, plasma $\beta$ maps for cool and hot regions were obtained using the temperature thresholds, e.g., on October 13 regions with temperature $T$ [$K$] < 5280 were defined as cool plasma regions and regions with $T$ [$K$] > 5280 represent regions with hot plasma. On October 17 regions with $T$ [$K$] < 5330 and $T$ [$K$] > 5330 are defined as cool and hot plasma regions, respectively (last two panels on Figure 2). Table 3 summarized the main findings.

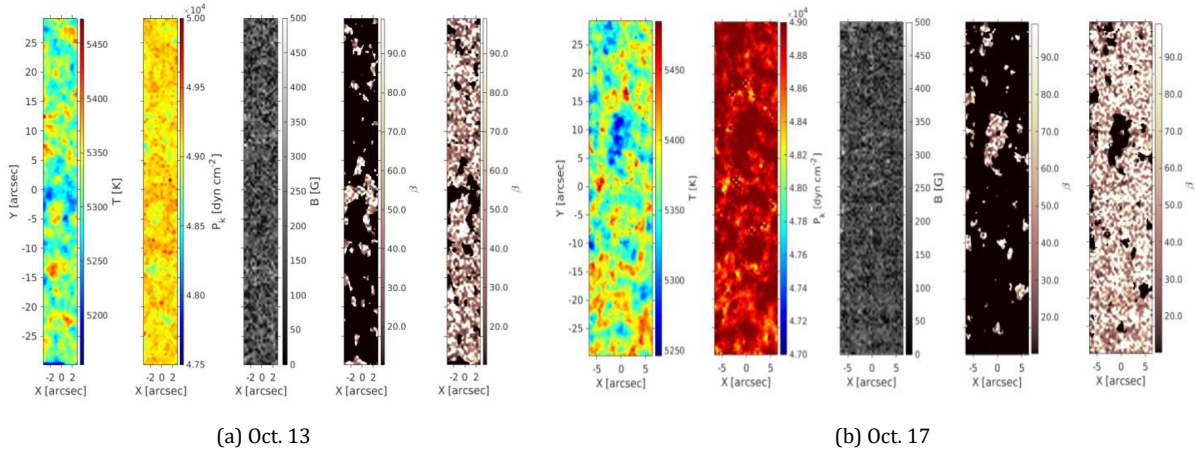

(a) Oct. 13                                                                 (b) Oct. 17

**Figure 2.** Left to right: Temperature ($T$ [$K$]), kinetic pressure ($P_k$ [$dyn\ cm^{-2}$]), magnetic field ($B$ [$G$]) from SIR code inversions at log $\tau$ = −1.0 and plasma $\beta$ maps for cool and hot regions. (a) October 13, 2019 (FOV 6.4″ × 59.6″) and (b) October 17, 2019 (FOV 13.2″ × 59.6″).

**Table 3.** Plasma $\beta$ in hot and cool plasma regions in the solar photosphere using the SIR code (GRIS) at optical depth $log\ \tau$ = −1.0 (Figure 2) using the Full Of View (FOV).

| Day | Region | B [G] | Plasma $\beta$ |
|---------|--------|--------|--------|
| Oct. 13 | Hot | 129.93 | 34.15 |
| Oct. 13 | Cool | 23.94 | 6.32 |
| Oct. 17 | Hot | 124.34 | 43.71 |
| Oct. 17 | Cool | 14.45 | 3.85 |

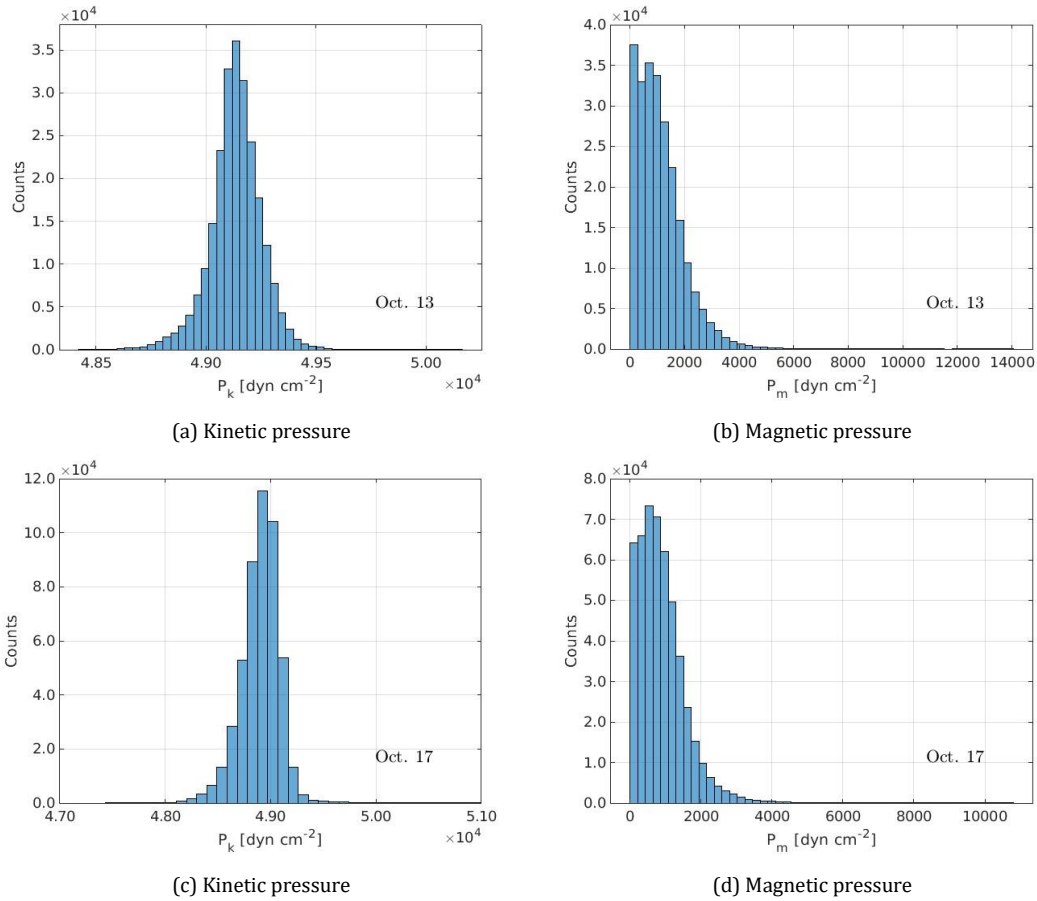

**Figure 3.** Distribution of the kinetic and magnetic pressure from SIR code (GRIS data) on Oct. 13 and Oct. 17, 2019.

Plasma $\beta$ is defined as the ratio of the kinetic pressure over the magnetic pressure (Bellot Rubio et al. (2020)-page 58). Thus, an analysis of the kinetic and magnetic pressure variations was performed using SIR inversions (GRIS data). Figure 3 shows the kinetic and magnetic pressure distribution from SIR code on Oct. 13 and 17. The kinetic pressure shows a Gaussian distribution with a mean value of $4.91 \times 10^4$ and $4.89 \times 10^4$ $dyn\ cm^{-2}$ on Oct. 13 and 17, respectively. While the magnetic pressure reveals a non-Gaussian heavy-tailed distribution (Rodríguez Gómez et al. (2020) and references therein). The mean magnetic pressure value corresponds to $1.10 \times 10^3$ $dyn\ cm^{-2}$ on October 13 and $888.92 \times 10^3$ $dyn\ cm^{-2}$ on October 17, 2019. Thus, the mean plasma $\beta$ value corresponds to 44.43 and 55.02 on October 13 and 17, respectively.

Additionally, the plasma $\beta$ was calculated in different optical depths as $\log \tau = -1.0, -1.5$ and $-2.0$, on October 13 and 17, 2019 using the complete FOV (Table 4). On both days, it is possible to see how the plasma $\beta$ decreases with the optical depth, from $\log \tau = -1.0$ to $\log \tau = -2.0$.

**Table 4. Plasma $\beta$ at different optical depths.**

| Day | Plasma $\beta$ $\log \tau = -1.0$ | Plasma $\beta$ $\log \tau = -1.5$ | Plasma $\beta$ $\log \tau = -2.0$ |
|---------|---------|---------|---------|
| Oct. 13 | 44.43 | 43.47 | 40.61 |
| Oct. 17 | 55.03 | 51.60 | 48.58 |

### 4.2. *Chromosphere*

We obtain chromospheric magnetic field estimates using the GRIS datasets, specifically applying inversion of the He I 10830 A triplet using the HAZEL code (section 3.2). Figure 4 shows the magnetic field derived from the HAZEL inversions on October 13 from 08:55:21 to 09:00:38 UT with a mean value $\bar{B}$ = 175.09 ± 72.11 $G$ and on October 17 between 09:48:45 and 09:59:16 UT with a magnetic field mean value of $\bar{B}$ = 80.20 ± 44.93 $G$.

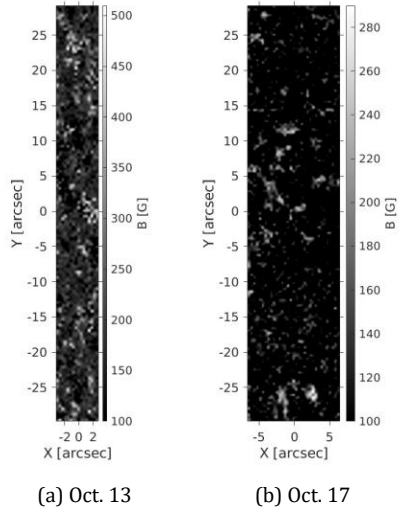

(a) Oct. 13              (b) Oct. 17

**Figure 4.** Magnetic field from He I 10830 A inversion using the HAZEL code. (a) Oct. 13, 2019 from 08:55:21 to 09:00:38 UT, (b) Oct. 17, 2019 from 09:48:45 to 09:59:16 UT.

However, to obtain plasma $\beta$, also density and temperature estimates in the chromosphere are necessary. For this purpose, we used the FAL-C model (Fontenla et al. 1993). This model includes a detailed description of particle diffusion, local ionization equilibrium and the transport of ionization energy for Helium emission as well as a constant abundance model. We highlight that the height formation of He I 10830 A has been under debate for many decades. However, in this work we used the height formation defined by Avrett et al. (1994b) as $h \sim 2100\ km$ and a temperature $T \sim 10^4\ K$. Using these values, it is possible to select the temperature and density interval from the FAL-C model. The temperature was selected from 10550 $K$ to 11150 $K$ and density between 3.81 × 10$^{10}$ and 4.06 × 10$^{10}$ $cm^{-3}$. We used a mean temperature $T$ = 10850 $K$ and mean density $N_e$ = 3.93 × 10$^{10}$ $cm^{-3}$. Additionally, a threshold to select magnetic field between 5 − 50 $G$ was applied. The plasma $\beta$ in the chromosphere obtained in this way are $\beta$ = 0.005 ± 0.007 on October 13 and $\beta$ = 0.006 ± 0.001 on October 17 (mean ± standard deviation).

### 4.3. *Transition Region*

Density in the transition region was obtained using the intensity ratios method described in Sect. 3.3. We obtain the density on October 19 using O IV intensity ratios from IRIS. The raster 23 (11:55:00 UT to 11:59:04 UT) was selected because it coincides with the Hinode/EIS period observations (Figure 5).

The intensity was obtained by integrating the total counts (DN) over the line profiles (Polito et al. 2016a) and converting them to physical units (phot s$^{-1}$arcsec$^{-2}$cm$^{-2}$) using the IRIS technical note 26 [4] and Lacatus et al. (2017).

---

[4] https://iris.lmsal.com/itn26/calibration.html

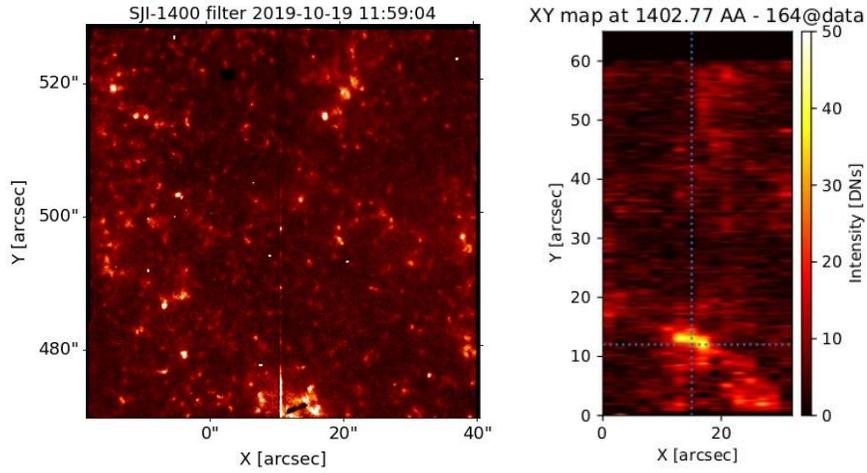

**Figure 5.**[3] Left to right: IRIS slit jaw image at 1400 A and raster map at 1402 A.

Density estimates were obtained using the IRIS_iris_ne routine. This routine requires the theoretical density-ratio of the 1399.77 A and 1401.16 A lines from the CHIANTI atomic database 8.0 [4]. This analysis was performed on the region shown on the raster map (Left panel, Figure 5), x-axis is between 5″ and 10″ and y-axis from 470″ to 480″. We were able to retrieve the 1399.77 A and 1401.16 A line intensities in the pixel, which corresponds to the position (8″, 475″). On the other pixels, these lines are weak or/and difficult to detect.

**Table 5.** Transition Region plasma $\beta$ estimations. Density derived from IRIS 1399.77 A and 1401.16 A lines, temperature from Rao et al. (2022) and the magnetic field from PFSS as the maximum value at a height of 1.014 $R_\odot$.

| Density $\log_{10} N_e [cm^{-3}]$ | Temperature $\log_{10} T [K]$ | B $[G]$ | Plasma $\beta$ |
|---|---|---|---|
| 9.80 | 4.92 | 12.01 | 0.005 |

To estimate the temperature using IRIS is possible by considering the ratio of the SIV 1406.1 A and 1401.16 A lines (Rao et al. 2022). However, a limitation from IRIS observations is that the S IV 1406.1 A line is at the edge of the spectral window, and in our case, is not measurable. For this reason we used the mean value from the values obtained by Rao et al. (2022) for quiet sun regions. Additionally, we used the Potential Field Source Surface (PFSS) model (Schrijver 2001; Schrijver & De Rosa 2003) to retrieve the magnetic field. To select the IRIS region in the PFSS model we convert the coordinates using SunPy[5]. The transition region is a very thin layer placed between chromosphere and the solar corona. Thus, we considered the transition region height at the coronal base 1.014 $R_\odot$ (Warmuth & Mann 2005). Table 5 summarizes the plasma estimates on the transition region.

## 4. *Corona*

We used data from Hinode/EIS and SDO/AIA to derive plasma $\beta$ through the solar corona. We observed a coronal hole on October 17 and a quiet sun region on October 19. We used Hinode/EIS observations on October 17 from 11:29:21 to 12:27:56 UT and October 19 from 11:00:21 to 11:58:57 UT to perform density diagnostic, with the Fe XII 186 A and Fe XII 195 A lines (Figure 6). The Fe XII 186 A and Fe XII 195 A lines are in the same spectral window and well separated. Thus, it is possible to fit them with two independent Gaussian. The Fe XII 186 A is blended by Ni XI 186.98 A in the red wing. The spectral points which correspond to the blend were not considered during fitting.

On the other hand, Fe XII 195 A is the strongest line, two Gaussian were used to fit this spectral line separated by 0.06 A during the fitting procedure. The densities are derived using CHIANTI atomic database from the line pair Fe XII 186/195 (section 3.4). A spatial binning of 4 × 4 pixels and stacked 2 min observations were performed. On Oct

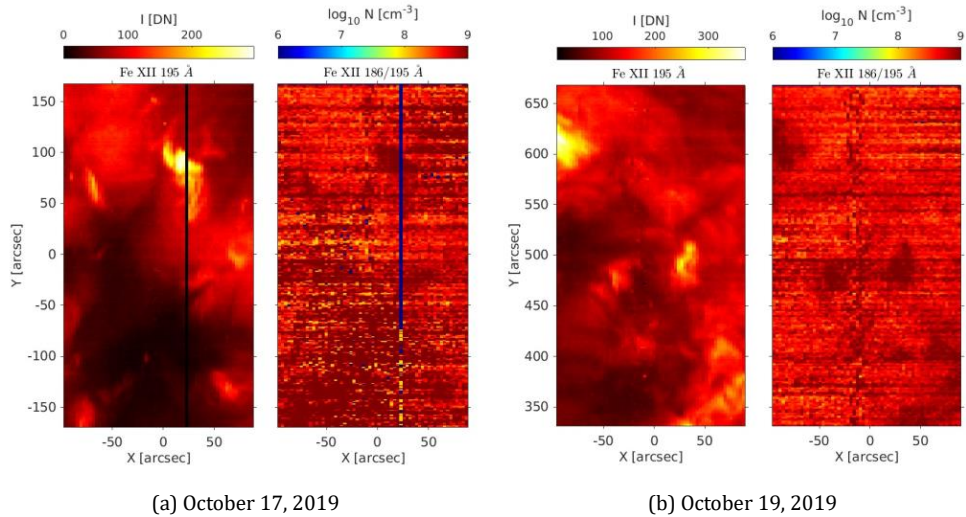

(a) October 17, 2019                      (b) October 19, 2019

**Figure 6.** Hinode/EIS Fe XII 195 A intensity maps and density maps derived from Fe XII 186/195 A line ratio on October 17 (a) and October 19 (b), 2019.

17 we observed a coronal hole and its signal was too weak specially for the Fe XII 186 A line, which strongly affects the quality of the derived density map. A quiet sun region was observed on Oct 19, and the signal from both lines was better as compared to October 17. Additionally, there is a vertical band of bad data due to a large number of cosmic rays during that time, especially on October 17.

We obtained the mean density value considering the complete FOV which resulted in $\log_{10} N_e = 8.92$ cm$^{-3}$ for October 17 and $\log_{10} N_e = 8.68$ cm$^{-3}$ for the quiet sun region on October 19. Additionally, we selected the coronal hole region on October 17, applying a threshold in the Fe XII 195 A intensity (I< 40 *DN*). We found that the density values inside that region have a mean value of $N_e = 10^{8.77}$ cm$^{-3}$. Thus, with the current datasets we were able to obtain density estimates but not temperature. Therefore, we used the reported temperature values, e.g., $\log_{10}(T) = 6.20$ in quiet sun obtained by Doschek & Warren (2019) and for the coronal hole case we used the temperature values ($\log_{10}(T) \approx 5.9 − 6.2$ *K*) presented by Zhu et al. (2022). They are derived from different intensity ratios observed by Hinode/EIS. The density and temperature obtained from Hinode/EIS can be located at heights below $1.2R_\odot$ (Goryaev et al. 2014). Thus, the magnetic field was obtained from PFSS between $1.03R_\odot$ and $1.2R_\odot$.

Additionally, we derived DEMs from the six SDO/AIA EUV filters on data from October 17, 2019 from 11:27:11 to 11:29:23 UT and October 19 from 11:01:59 to 11:03:47 UT. We coaligned SDO/AIA at 211 A with Hinode/EIS Fe XII 195 A intensity maps before performing the DEMs analysis. For better signal-to-noise ratio for the DEM analysis, we added for each AIA filter 10 subsequent images (corresponding to 2 minutes) and applied a spatial binning of 4 × 4 pixels. To obtain density we used Equation 9, for the coronal hole case on October 17 (Figure 7), we used $\mu$ = 0.64, $m_H$ = 1.67 × 10$^{-27}$ kg, $T$ = 0.9 MK, and we obtain a scale height $h$ = 42 Mm (Saqri et al. 2020). For the quiet sun region observed on October 19 we used quiet sun temperature ∼ 1.5 MK (Morgan & Taroyan 2017) which correspond to a scale height for quiet sun to $h$ = 70.5 Mm (Figure 8).

A coronal Hole (CH) was observed using Hinode/EIS and SDO/AIA on October 17, 2019. To select the CH region in SDO/AIA data (211A), we applied an intensity threshold (I< 2500 $DN$). The mean density obtained from DEMs analysis corresponds to $N_e$ = 6.38×10$^8$ $cm^{-3}$, and the mean temperature value is 1.23×10$^6$ $K$. A quiet sun region was observed on October 19 using both instruments. The mean density and temperature obtained from DEMs correspond to $N_e$ = 7.43×10$^8$ $cm^{-3}$ and 1.29×10$^6$ $K$, respectively. In this case we observe against the disk, and considering that most of the emission comes from the base of the corona. Density and temperature estimations using Hinode/EIS and SDO/AIA in the coronal hole area show similar values than the quiet sun region. Also, the temperature obtained from SDO/AIA data in the coronal hole region shows a lower value compared to the quiet Sun area in both days (Table 6).

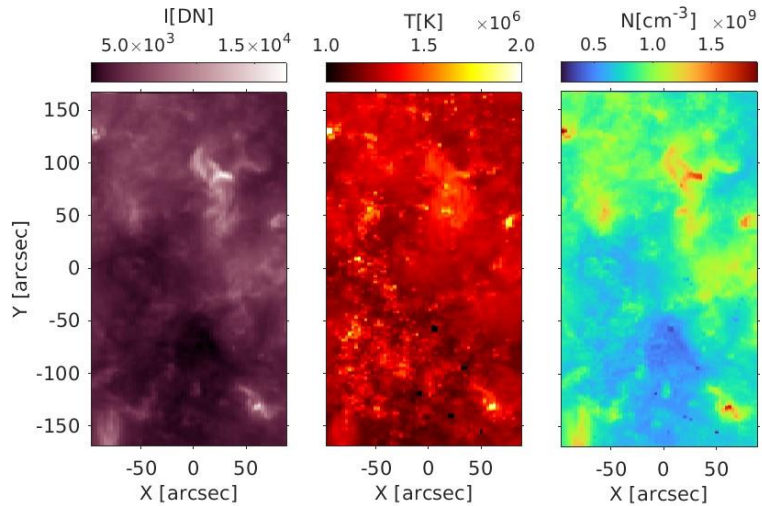

**Figure 7.** Left to right: AIA intensity maps at 211 A on October 17, 2019, and corresponding temperature and density maps obtained from the DEM analysis.

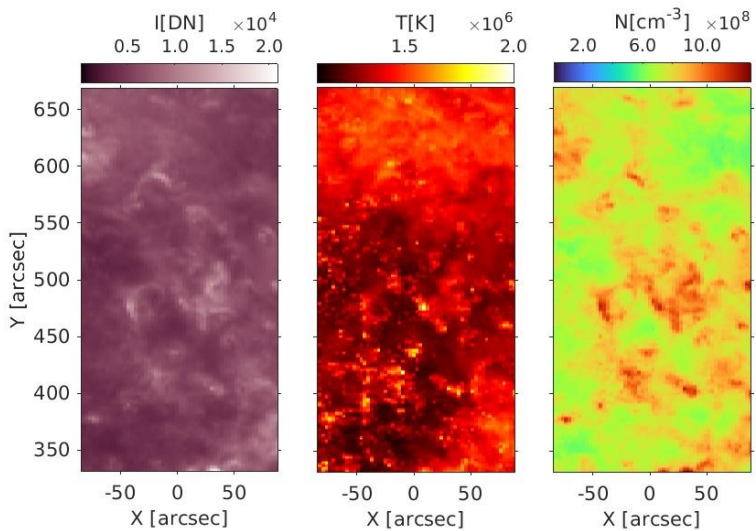

**Figure 8.** Left to right: AIA intensity maps at 211 A on October 19, 2019, and corresponding temperature and density maps obtained from the DEM analysis.

We note that the temperature values derived for the coronal hole region may be an overestimate, as the coronal hole is small and therefore the stray light effect from the surrounding regions of much higher intensity is expected to be substantial (cf. Saqri et al. (2020)).

## 5. DISCUSSION AND CONCLUSIONS

We obtain the plasma $\beta$ in quiet sun regions using data from a joint campaign of several space-borne and ground based observatories. The main aim was to have observations simultaneously and retrieve a complete picture of plasma $\beta$ through the different layers of solar atmosphere. The photosphere and chromosphere were observed by GREGOR Solar telescope, while the transition region and the corona was observed by different spacecraft (IRIS, Hinode/EIS, SDO/AIA). Figure 9 shows the derived observational values for the plasma beta on top of the profiles from Gary (2001) derived from Rodríguez Gómez et al. (2019). The results of this study can be summarized as follows:

**Photosphere**

Table 6. Plasma $\beta$ at coronal heights: in the Coronal Hole region (CH) and quiet sun region (QS).

| Date | Data | Feature | Density | Temperature | B | Plasma $\beta$ |
|------|------|---------|---------|-------------|---|----------------|
|      |      |         | $log_{10} \, N_e \, [cm^{-3}]$ | $log_{10} \, T \, [K]$ | $[G]$ | |
| Oct. 17 | EIS | CH | 8.63 | 6.05 | 1.85 | 2.02 |
| Oct. 17 | EIS | QS | 8.78 | 6.20 | 2.13 | 1.52 |
| Oct. 19 | EIS | QS | 8.68 | 6.20 | 2.47 | 1.14 |
| Date | Data | Feature | Density | Temperature | B | Plasma $\beta$ |
|      |      |         | $N_e \, [cm^{-3}]$ | $T \, [K]$ | $[G]$ | |
| Oct. 17 | AIA | CH | $6.38 \times 10^8$ | $1.23 \times 10^6$ | 1.85 | 1.59 |
| Oct. 17 | AIA | QS | $9.01 \times 10^8$ | $1.29 \times 10^6$ | 2.13 | 1.77 |
| Oct. 19 | AIA | QS | $7.43 \times 10^8$ | $1.29 \times 10^6$ | 2.47 | 1.09 |

- We retrieve plasma $\beta$ values in the solar photosphere mainly using ground-based observations from GREGOR solar telescope.

- Temperature estimation from HiFI Speckle reconstructed images show values as expected from theoretical values in the blue continuum and G-Band. These results demonstrate that this technique is reliable in retrieving temperature in photosphere. We used those temperature values together with the magnetic field from HMI/SDO and density estimations from FALC model to obtain plasma $\beta$ values. These values lie inside the expected values in the photosphere (cf. Figure 9).

- To analyze plasma $\beta$ from SIR code (GRIS data), we explored the kinetic and magnetic pressure variations in optical depth $log\tau = -1.0$. The kinetic pressure shows a Gaussian distribution, while the magnetic pressure reveals a non-Gaussian heavy-tailed distribution. This kind of distribution shows the complexity of the magnetic field regimes in the quiet sun obtained from high-resolution observations. As well as hot and cool plasma regions using FOV at the same optical depth. We observed that hot regions show high plasma $\beta$ compared to the coolest ones. Implying, that the plasma $\beta$ in those regions is controlled mainly by the kinetic pressure.

- We explored the plasma $\beta$ at different optical depths (log $\tau = -1.0, -1.5$ and $-2.0$) obtained from the SIR code. These values decrease with the optical depth, e.g., lower values were found at optical depths log $\tau = -1.5$ and $-2.0$ compared to log $\tau = -1.0$. Unfortunately, these values cannot be plotted over Figure 3 of (Gary 2001), because the correspondence between optical depth and height of the radiation is not known ($R_\odot$). Some methods were developed to find the relationship between optical depth and height in the solar photosphere but their relation is not clear yet, e.g., Borrero et al. (2019).

- In general, plasma $\beta$ values retrieved from both instruments are inside the expected values in the photosphere (Figure 9) $\beta \approx 0.2 \, to \, 10^2$.

**Chromosphere**

- We were able to retrieve the magnetic field in the solar chromosphere using GRIS-GREGOR He I 10830 A data and the HAZEL code. The derived magnetic field mean values range between 80 $G$ and 175 $G$. In this work, we had limited datasets to describe the chromosphere. For this reason, we used FALC model to retrieve information related to density and temperature at the He I 10830 A formation height.

- We used those values to obtain plasma $\beta$ values in the chromosphere. As result, we obtain values that agree with the expected values in the chromosphere. We observe that those values are lower than the photospheric values. It can be explained because in the solar chromosphere, the magnetic pressure dominates the kinetic pressure, explaining the lower plasma $\beta$ values in the chromosphere.

**Transition Region**

- We obtain the plasma $\beta$ in the transition region using the IRIS dataset and the magnetic field from the Potential Field Source Surface (PFSS).
- The OIV lines observed by IRIS are frequently used to obtain densities. Specifically, 1399.77 A, 1401.16 A and 1404.8 A, but they are very weak in quiet sun regions. Other lines like S IV 1404.8 A and 1406.1 A can be used to retrieve density estimations, but they are not available in this dataset. Thus, we used O IV 1399.77 A and 1401.16 A to obtain density estimates in the transition region. We selected the pixel with the highest intensity on October 19 (Figure 5). The density value corresponds to $N_e = 10^{9.8} cm^{-3}$. This value agrees with the quiet sun estimations obtained in previous works, e.g., Rao et al. (2022) and Polito et al. (2016a). Using the temperature value obtained by Rao et al. (2022) and the magnetic field retrieved from PFSS, we obtain a plasma $\beta$ value of 0.005. It is in agreement with the expected limits in transition region.

- The magnetic field value obtained from PFSS at 1.014 $R_\odot$ agrees with values obtained by Trujillo Bueno et al. (2011). They used 1D models of the solar atmosphere, based on the Q/I and U/I line center signals. Through the Hanle effect, these lines are sensitive to magnetic fields at the base of the solar corona, with values ranging between 10 and 100 G (Trujillo Bueno & del Pino Alemán 2022). It is expected that in the future observational data help to accurately describe the solar magnetic field in the transition region. To understand its impact on the solar corona dynamics.

**Corona**

- We retrieve the plasma $\beta$ values in the solar corona from 1.03 $R_\odot$ to 1.20 $R_\odot$ using density and temperature estimations from Hinode/EIS and SDO/AIA datasets, and the coronal magnetic field retrieved by PFSS.

- Density estimations for the quiet sun corona are in agreement with the values obtained from the CODET model (Rodríguez Gómez et al. 2018) and values obtained through Fe XII lines reported by Zhu et al. (2022); Long et al. (2013); Young et al. (2009). The Coronal Hole magnetic field from PFSS is in agreement with the values presented by Linker et al. (2021).

- We obtained plasma $\beta$ values from Hinode/EIS spectroscopy and SDO/AIA DEM diagnostics in the quiet sun and in the coronal hole region. In general, density and temperature values in the CH region and quiet sun are similar. The plasma $\beta$ value on October 17 related to the CH region appears in the limit defined by Gary (2001). Some factors can affect the coronal hole characterization, e.g., the stray light effect from the surrounding CH regions that can affect the temperature estimate. Furthermore, the coronal hole magnetic field estimations, e.g., the magnetic field obtained from PFSS extrapolations. Additionally, Rodríguez Gómez et al. (2019) reported the same behavior at the same coronal heights. It can imply that the limits of plasma $\beta$ at that coronal height need to be enlarged.

Figure 9 summarizes our findings through the solar atmosphere in the quiet sun regime. We over plotted our results on the adapted Figure 3 of Gary (2001) previously presented by Rodríguez Gómez et al. (2019). We defined different intervals related to the different atmospheric layers, e.g., the transition region observed by IRIS was defined at the height corresponding to the coronal base at 1.014 $R_\odot$ (red dotted line). The solar corona was observed using two different spacecraft: Hinode/EIS and SDO/AIA covering the coronal heights from 1.03 $R_\odot$ to 1.20 $R_\odot$ (blue dotted lines). Quiet sun regions were observed on October 13, 17 and 19 and plotted with a blue circle. The plasma $\beta$ calculated from the Coronal Hole region is denoted with the red circle. In coronal heights, we plot quiet and coronal hole regions derived from SDO/AIA using soft colors. On October 13 we could observe uniquely using the ground based telescopes. Thus, our findings on that day reflect only plasma $\beta$ in the photosphere and chromosphere. In the solar photosphere, we plotted the mean value from HiFI and GRIS datasets.

In general, we can describe the plasma $\beta$ from the photosphere to solar corona quite well with the current datasets. However, a wide study in the transition region and chromosphere is necessary because most of the coronal problems that the solar community is trying to solve, needs a clear plasma description at those heights. The data from SPICE onboard Solar Orbiter (launched in 2020) will provide detailed information about the plasma on the chromosphere, transition region and corona, observed far-side and at different heliocentric distances.

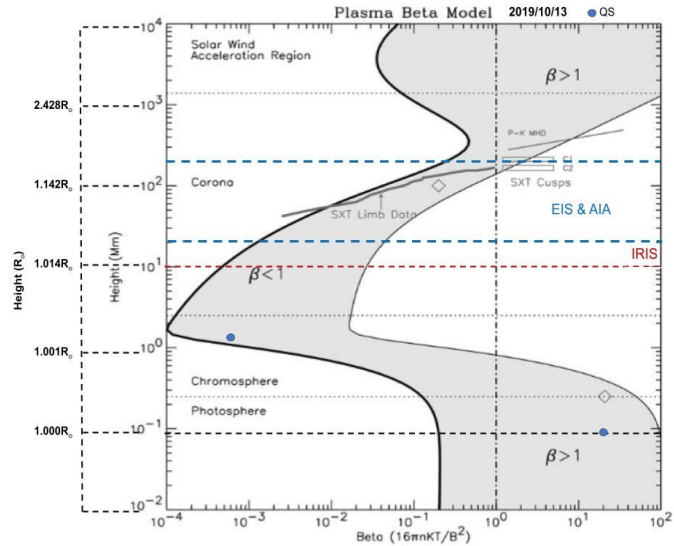

(a) Oct. 13

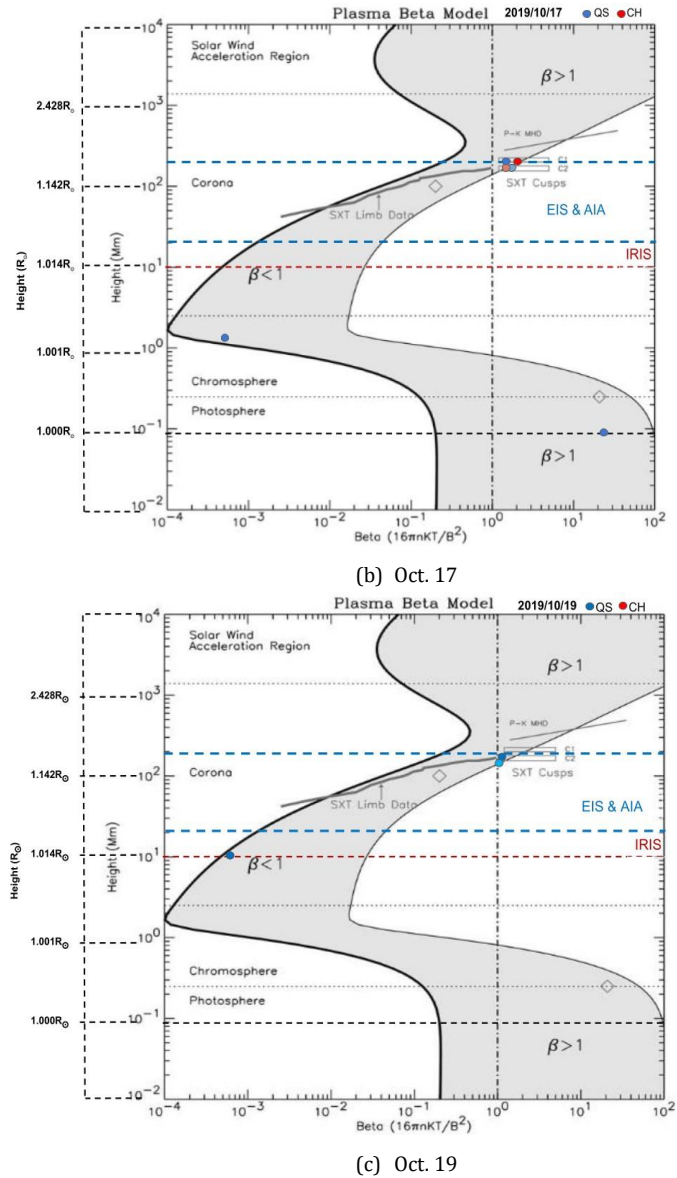

**Figure 9.** Plasma $\beta$ through the solar atmosphere on Oct. 13 (a), Oct. 17 (b) and Oct. 19 (c), from different ground-based and space-based data, photosphere-chromosphere: GREGOR HiFI and GRIS, Transition region: IRIS (red dotted line), corona: Hinode/EIS interval and SDO/AIA denoted by the blue dotted line.

J.M.R.G. thanks Judith Palacios for her support during the observation campaign at GREGOR solar telescope and Peter Young for his help and productive discussions about IRIS and EIS data analysis. This work was supported by NASA Goddard Space Flight Center through Cooperative Agreement 80NSSC21M0180 to Catholic University, Partnership for Heliophysics and Space Environment Research (PHaSER) and the European Union's Horizon 2020 research and innovation program under grant agreement No. 824135 (SOLARNET). C.K. acknowledges funding from the European Union's Horizon 2020 research and innovation programme under the Marie Skl odowska-Curie grant agreement No 895955. PG acknowledges support of the project VEGA 2/0048/20. SJGM is grateful for the support of the European Research Council through the grant ERC-2017-CoG771310-PI2FA, the MCIN/AEI/ 10.13039/501100011033 and "ERDF A way of making Europe" through grant PGC2018-095832-B-I00, and the project VEGA 2/0048/20. The 1.5-meter GREGOR solar telescope was built by a German consortium under the leadership of the Leibniz-Institut für Sonnenphysik in Freiburg with the Leibniz-Institut für Astrophysik Potsdam, the Institut für Astrophysik Göttingen, and the Max-Planck-Institut für Sonnensystemforschung in Göttingen as partners, and with contributions by the Instituto de Astrofísica de Canarias and the Astronomical Institute of the Academy of Sciences of the Czech Republic. The redesign of the GREGOR AO and instrument distribution optics was carried out by KIS whose technical staff is gratefully acknowledged.